\documentclass[10pt,twocolumn,letterpaper,superscriptaddress]{revtex4}

\usepackage[paperwidth=215mm,paperheight=310mm,centering,hmargin=1.5cm,vmargin=2cm]{geometry}
\usepackage{multibbl}
\usepackage[pdftex]{graphicx}
\usepackage{amsmath,SIunits}
\usepackage[latin1]{inputenc}
\usepackage{latexsym,stmaryrd}
\usepackage[sort&compress]{natbib}
\usepackage[pdftex]{graphicx}
\usepackage{color}
\bibpunct{[}{]}{,}{n}{}{,}

\makeatletter
\newcommand*{\citenst}[2][]{%
  \begingroup
  \let\NAT@mbox=\mbox
  \let\@cite\NAT@citenum
  \let\NAT@space\NAT@spacechar
  \let\NAT@super@kern\relax
  \renewcommand\NAT@open{[}%
  \renewcommand\NAT@close{]}%
  \citep{#2}%
  \endgroup
}
\makeatother

\begin{document}

\title{Anderson photon-phonon co-localization in certain random superlattices}

\author{G. Arregui}
\email{guillermo.arregui@icn2.cat}
\affiliation{Catalan Institute of Nanoscience and Nanotechnology (ICN2), CSIC and BIST, Campus UAB, Bellaterra, 08193 Barcelona, Spain}
\affiliation{Dept. de F\'{i}sica, Universitat Autonoma de Barcelona, 08193 Bellaterra, Spain}
\author{N. D. Lanzillotti-Kimura}
\affiliation{Centre de Nanosciences et de Nanotechnologies, CNRS, Univ. Paris-Sud, Universit\'e Paris-Saclay, C2N - Marcoussis, 91460 Marcoussis, France}
\author{C. M. Sotomayor-Torres}
\affiliation{Catalan Institute of Nanoscience and Nanotechnology (ICN2), CSIC and BIST, Campus UAB, Bellaterra, 08193 Barcelona, Spain}
\affiliation{ICREA - Instituci\'o Catalana de Recerca i Estudis Avan\c{c}ats, 08010 Barcelona, Spain}
\author{P. D. Garc\'{i}a}
\affiliation{Catalan Institute of Nanoscience and Nanotechnology (ICN2), CSIC and BIST, Campus UAB, Bellaterra, 08193 Barcelona, Spain}

\date{\today}

\small

\begin{abstract}
Fundamental concepts in quantum physics and technological applications ranging from the detection of gravitational waves to the generation of stimulated Brillouin scattering rely on the interaction between the optical and the mechanical degrees of freedom.\ A key parameter to engineer this interaction is the spatial overlap between the two fields, optimized in carefully designed resonators on a case-by-case basis.\ Disorder is an alternative strategy to confine light and sound at the nanoscale.\ However, it lacks an \textit{a priori} mechanism guaranteeing a high degree of co-localization due to the inherently complex nature of the underlying interference process.\ Here, we propose a way to address this challenge by using GaAs/AlAs vertical distributed Bragg reflectors with embedded geometrical disorder.\ Due to a remarkable coincidence in the physical parameters governing light and motion propagation in these two materials, the equations for both longitudinal acoustic waves in the growth direction and normal-incidence light become practically equivalent for excitations of the same wavelength.\ This guarantees spatial overlap between photons and phonons leading to a statistically significant enhancement in the vacuum optomechanical coupling rate, $g_{0}$, and making this system a promising candidate to explore Anderson localization of high frequency ($\sim$ 20 GHz) phonons enabled by cavity optomechanics.\
\end{abstract}


\maketitle

The interaction between electromagnetic radiation and mechanical motion through radiation pressure, photoelasticity or thermoelasticity~\cite{arcizetmicromirror,ruellophotothermal,eichenfieldomcrystals} in optomechanical systems spans at least twenty orders of magnitude in mass and ten in frequency \cite{reviewoptomechanics}.\ These systems are suited to test fundamental quantum physics with \textit{macroscopic} objects \cite{verhagensplitting,oconnelmicrowave} and are promising candidates as transducers in quantum systems~\cite{groblacherroomT}, ultra-high sensitivity mass and force sensors~\cite{gavartinnanotech}, high bandwidth accelerometers~\cite{krauseaccelerometer}, optical delay-lines~\cite{OITScience}, high-tunability optical filters~\cite{OpticalFilterPainter} and wavelength conversion~\cite{WavelengthConversionHill}.\ By designing and optimizing a defect in a periodically patterned slab structure~\cite{yablonovitch,acousticbandstructure}, it is possible to achieve photon and phonon confinement with control of both the coupling efficiency and the dissipation channels~\cite{eichenfieldomcrystals,SafaviDesign2D} using a silicon-compatible technology~\cite{CMOSFoundry}.\ Nevertheless, unavoidable fabrication imperfections open undesirable leakage channels for photons and phonons that degrade both the mechanical and optical quality (Q-)factors~\cite{minkovdisorder}, leading to a reduced optomechanical interaction and limiting the ambition of design efforts.\ Further reducing fabrication disorder~\cite{improvingfab} or minimizing its effect~\cite{topologicallyprotected} have been the main approaches to circumvent this issue.\ However, disorder can be used to confine light and mechanical motion within silicon nanobeams to explore cavity optomechanics~\cite{ALoptomechanicsnanobeam}.\ This approach also points to the use of optomechanical displacement read-out as a detection mechanism to probe the fundamental nature of Anderson localization of GHz phonons.\ Nevertheless, the electromagnetic and displacement fields interfere independently within the perturbed structure and hardly co-localize, thus hindering the maximum achievable value of the vacuum optomechanical coupling rate $g_{0}$.

 \begin{figure}[t!]
  \includegraphics[width=\columnwidth]{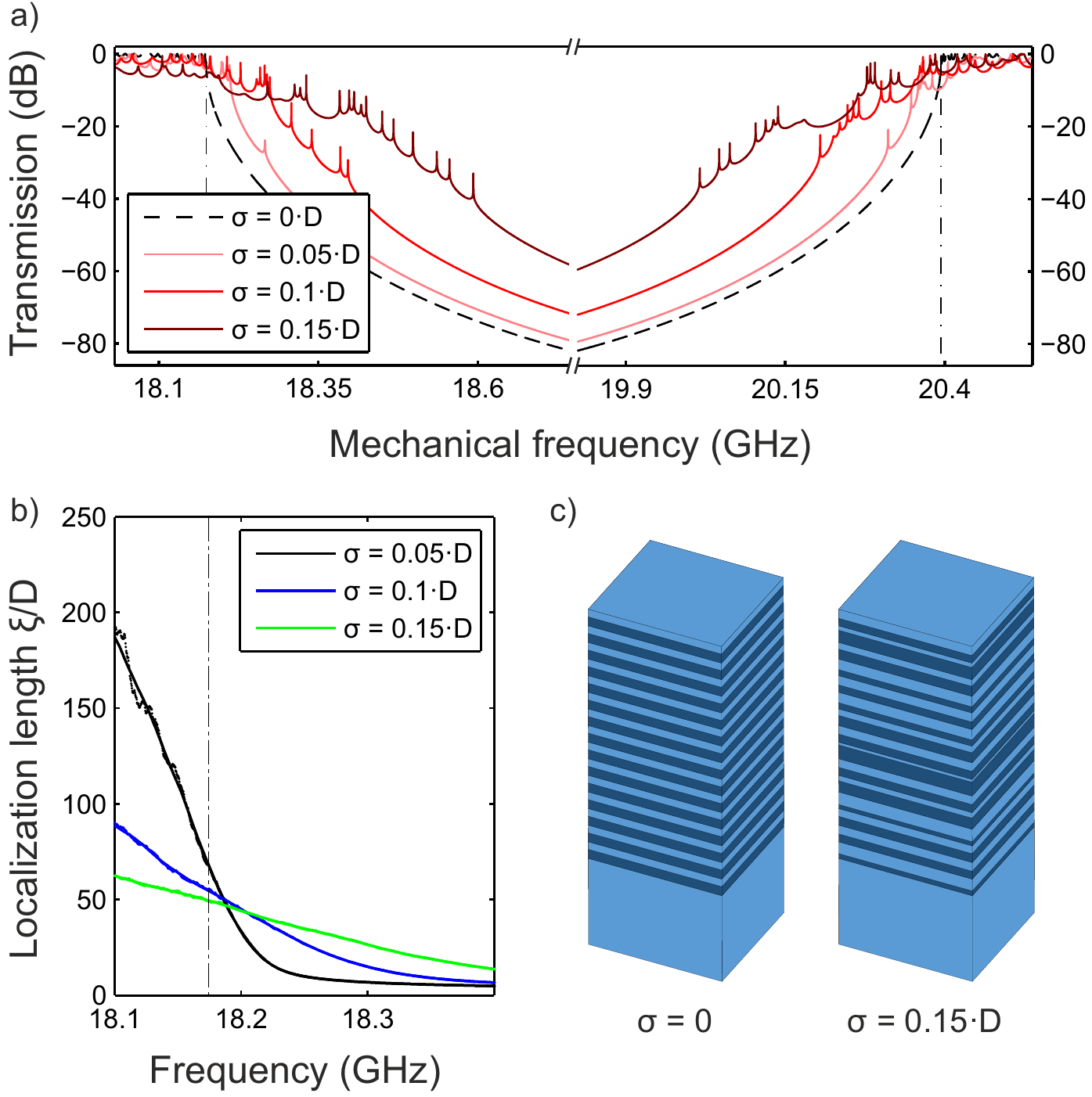}
      \caption{ \label{1} \textbf{Phonon localization in disordered GaAs/AlAs superlattices.} (a) Transmission spectrum of perturbed
      GaAs/AlAs superlattices ($d_{1}$=73.48 nm, $d_{2}$=61.88 nm, $N$=600), with Gaussian disorder $\mathcal{N}(0,\sigma^2)$ in the position of the interface inside the unit cell.\ (b) Localization length $\xi$
      normalized by the unit cell thickness $D$ for several standard deviations $\sigma$. The lines that serve as a guide to the eye are computed as an average over small frequency ranges.\
      (c) Schematic of a perfect superlattice ($\sigma = 0$) and a perturbed one ($\sigma = 0.15\cdot D$) on a thick GaAs substrate.}
    \end{figure}

Single-molecules~\cite{singlemoleculeoptomechanics}, high-stress silicon nitride membranes~\cite{groblacherroomT}, epitaxially-grown planar distributed Bragg reflectors (DBRs)~\cite{doublemagiccoincidence} or  micropillars~\cite{micropillarsfabrice,micropillarsBariloche}, are alternatives to study cavity optomechanics in regimes otherwise difficult to explore with optomechanical crystals fabricated with standard electron-beam lithography.\ High mechanical frequencies $\Omega_{m}$ and/or mechanical quality factors $Q_{m}$ could enable quantum coherent control of both the mechanical resonator and the light field without pre-cooling the thermal bath of the system, thus alleviating the disadvantages imposed by cryogenic temperatures.\ Molecular beam epitaxy (MBE) grown planar superlattices with different acoustic and optical impedances are an ideal platform for time-resolved room-temperature cavity optomechanics~\cite{micropillarsBariloche,SesinTuningDBR} with up to THz mechanical vibrations.\ In particular, the GaAs/AlAs Fabry-P\'{e}rot resonators proposed in Ref.~\cite{doublemagiccoincidence} are designed to confine photons optimally and behave simultaneously as resonators for acoustic phonons with the same wavelength, quality factor and field profile.\ Here, we show that any supperlattice composed by an arbitrary combination of these two materials exhibits almost the same mechanical and optical interference pattern even when induced by disorder.\ We propose, theoretically, to use GaAs/AlAs disordered multilayers as a cavity-optomechanical system to enable optomechanical phenomena in the Anderson-localization regime and to explore Anderson localization of coupled (photon-phonon) excitations~\cite{Marquardt}.\ Moreover, due to the optomechanical nature of the detection mechanism in time-resolved pump-probe experimental techniques~\cite{DLKBasics}, these systems provide an ideal platform to actually probe Anderson localization of mechanical vibrations in very high frequency regions so far unexplored.

\begin{figure}[t!]
  \includegraphics[width=\columnwidth]{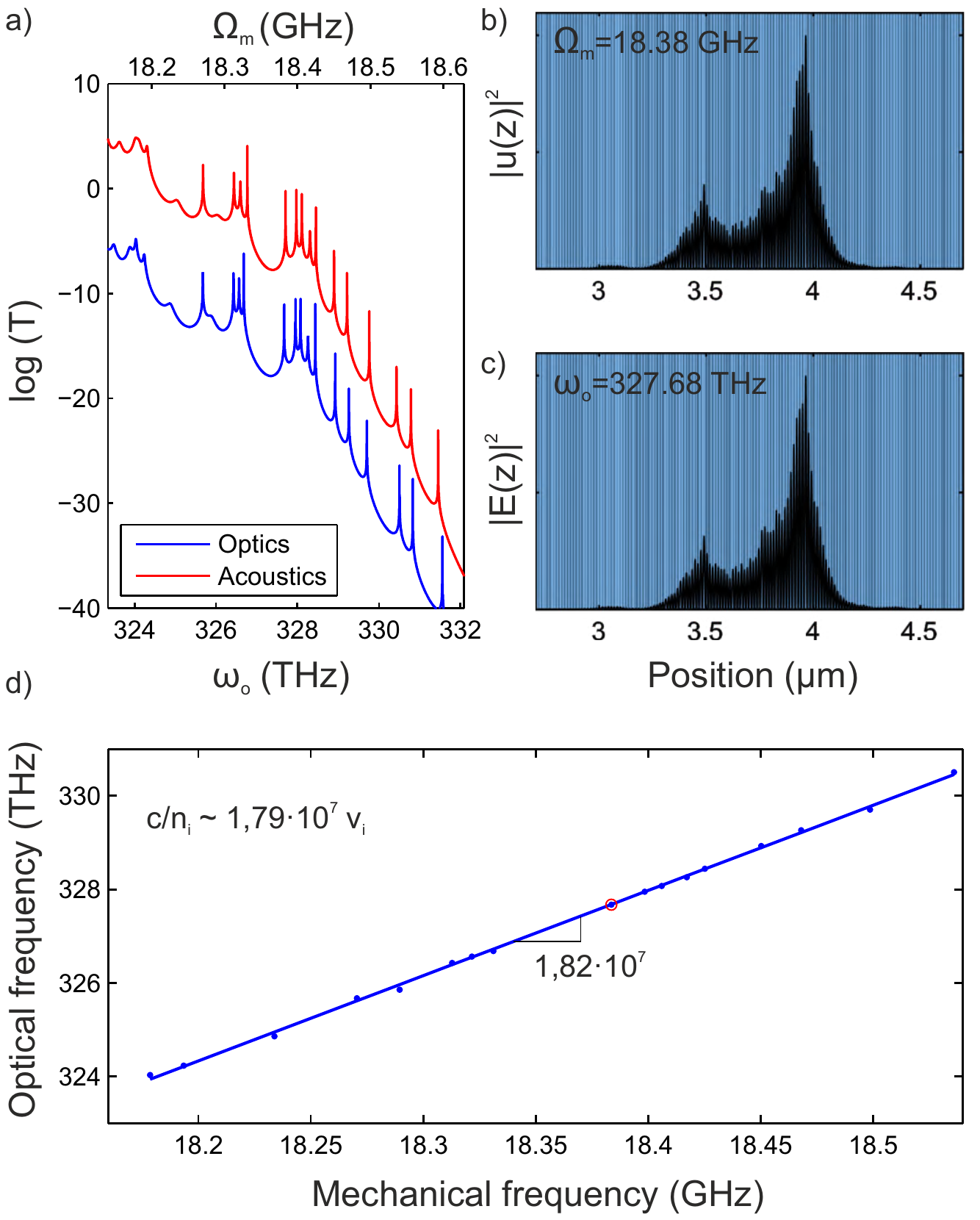}
    \caption{ \label{2} \textbf{Photon-phonon co-localization.} (a) Optical and acoustic transmission spectra of a disordered
    GaAs/AlAs superlattice with $d_{1}$=73.48 nm, $d_{2}$=61.88 nm, $N$=600 and $\sigma$=0.15$\cdot D$ at the low-frequency optical and mechanical band-gap edge, respectively.\ The energy density of the displacement field $u(z)$ and the electric field $E(z)$ of the highlighted pair in (d) with a red dot are plotted in (b) and (c), showing perfect co-localization.\ The resonant frequencies of all the mechanical and optical eigenmodes are mapped onto each other and plotted in (d).}
\end{figure}

To study the optical and acoustic properties of these superlattices, we use a transfer matrix formalism~\cite{rytovmodel} (see supplementary information).\ In this system, the solution to the continuum-mechanics equations in absence of sources and to the Maxwell's equations at normal incidence can be solved with exactly the same formalism~\cite{TMM}, provided that the acoustic impedances $Z_{i}$ are replaced by refractive indexes $n_{i}$ and the longitudinal acoustic velocities $v_{i}$ by the speed of light, $\frac{c}{n_{i}}$, in the different \emph{i} layers.\ Using standard values~\cite{adachi} for the optical and mechanical coefficients for optical photons and microwave phonons in AlAs (1) and GaAs (2), we have $n_{1}/n_{2} = 0.838 \sim 0.834 = Z_{1}/Z_{2}$ and $n_{1}/n_{2} = 0.838 \sim 0.844 = v_{2}/v_{1}$, which formally implies

\begin{equation}
\label{conditions}
\mathbf{M}_{ac}(\omega) \cong \mathbf{M}_{op}(K\omega),
\end{equation}
where $\mathbf{M}_{ac}(\omega)$ and $\mathbf{M}_{op}(K\omega)$ are the matrices connecting the acoustic and optical coefficients in the propagating/counter-propagating plane-wave basis at two consecutive layers and $K = \frac{c/n_{i}}{v_{i}}$ (details in the supplementary material).\ Any \emph{arbitrary} GaAs/AlAs superlattice exhibiting a mechanical eigenmode with a field profile $u(z)$ and frequency $\Omega_{m}$ will also support an optical eigenmode with a frequency $\omega_{o}\sim K\cdot\Omega_{m}$ and field profile $E(z)\sim u(z)$.\ To illustrate this, we use a DBR structure with a unit cell formed by an AlAs layer of thickness $d_{1}=73.48$ nm and a GaAs layer of thickness $d_{2}=61.88$ nm (technically known as a $\lambda/4, \lambda/4$ DBR).\ This structure opens simultaneously a maximum-width first-order band-gap with a center wavelength of 870 nm for optical photons and 19 GHz for microwave phonons.\ We fix GaAs for the substrate and the surface layers in the simulated structure and we introduce a zero-mean Gaussian disorder in the position of the interface between the two materials with varying standard deviation $\sigma$, whilst the period $D=d_{1}+d_{2}=135.36$ nm is kept constant.\ Fig.~\ref{1} shows the acoustic transmission spectrum of a disordered superlattice with $N=600$ periods for different disorder levels, $\sigma$, around the first-order odd band-gap.

Non-interacting electronic Bloch modes in atomic crystals undergo random multiple scattering in the presence of disorder, eventually leading to the (Anderson) localization of the wave function~\cite{absenceofdiffusion}.\ The electromagnetic and displacement fields in artificial photonic and phononic crystals are also sensitive to disorder, especially at the band-gap edges where the group velocity falls (ideally) to zero~\cite{sjohn}, equally giving rise to disorder-induced or Anderson localization.\ In a one-dimensional structure such as a GaAs/AlAs superlattice, the presence of Anderson-localized modes can be detected in the reflection/transmission spectrum through Lorentzian-shaped resonances centered near the corresponding eigenfrequencies with a free spectral range $\delta \omega$ larger than the mean linewidth $\Delta \omega$~\cite{thoulesscriterion}.\ Fig.~\ref{1}(a) shows these narrow mechanical resonances in the transmission spectra visibly satisfying such criteria and populating a spectral band that broadens with increasing disorder level~\cite{lifshitztail}.\ The ensemble-averaged decay of such localized modes occurs with a \emph{localization} length, $\xi$.\ It determines the minimum length of a finite sample for which it is possible to statistically observe these modes as $L \geq \xi$~\cite{shenglocalization}, where $L=N\cdot D$ is the total length of the structure.\ Fig.~\ref{1}(b) computes $\xi$ as a function of disorder level and frequency, using the scaling of the ensemble-averaged logarithmic transmission $<log(T)>$ $\propto$ $-\frac{L}{\xi}$ and shows that the system is deep in the localization regime ($\xi\ll L$) near the band edges for the range of disorder considered.

When the condition (\ref{conditions}) is fulfilled, the optical spectrum is almost identical to the mechanical one when plotted with scaled frequency, as shown in Fig.~\ref{2}(a).\ The field profiles of a pair of optical and mechanical modes with same \emph{scaled} frequency are also perfectly overlapping in real space, as shown in Fig.~\ref{2}(b) and (c), respectively.\ When the mechanical and optical resonant frequencies are mapped onto each other in spectral order,  we recover - Fig.~\ref{2}(d) - the predicted spectral behavior with $\omega_{o}\approx K\cdot\Omega_{m}$ and $K=1.82\cdot 10^7 \sim \frac{c/n_{i}}{v_{i}}$.\ These well-confined spatially-overlapping modes can interact with each other and are therefore candidates to explore optomechanical effects.\ As expected from confined mechanical and light modes, the deformation profile associated with a normal mode $u_{m}(z)$ will locally change the optical properties of the structure.\ Thus, the electromagnetic normal modes $E_{n}(z)$ will be affected, giving rise to an optical frequency shift (dispersive optomechanics) and a quality factor change (dissipative optomechanics).\ In GaAs/AlAs superlattices and in the frequency range of interest, two main acousto-optic interaction mechanisms need to be considered~\cite{ruello,wright}.\ First, the displacement of the $N+1$ boundaries, or moving boundary effect, will change the interference pattern of multiple light paths.\ Second, the photoelastic effect will induce a change in the bulk permittivity tensor $\mathbf{\epsilon}$ that can be written as the tensor product $d(\epsilon^{-1})_{ij}=P_{ijkl}U{kl}$, where $\mathbf{U}$ is the second-order strain tensor and $\mathbf{P}$ the fourth-order photoelastic tensor~\cite{matsuda}.\ To account for the frequency shift induced by these two mechanisms, first-order perturbation theory applied to Maxwell's equations~\cite{joannopoulosPT} in the multilayered system leads to the following expressions:
\begin{subequations}
\begin{align}
    \label{gmb}
    g_{mb}=-\frac{\omega_{o}}{2} \frac{\sum^{N+1}_{i=1} u_{m}(z_{i})(\epsilon_{i-1}-\epsilon_{i})\lvert E_{n}(z_{i})\rvert^2}{\int_{0}^{L}\epsilon(z)\lvert E_{n}(z)\rvert^2 dz} x_{zpf}
\end{align}
\begin{align}
\label{gpe}
    g_{pe}=-\frac{\omega_{o}}{2} \frac{\int^{0}_{L} P_{12}(z)\frac{\partial u_{m}}{\partial z}(z)\epsilon(z)^2 \lvert E_{n}(z)\rvert^2}{\int_{0}^{L}\epsilon(z)\lvert E_{n}(z)\rvert^2 dz} x_{zpf}
\end{align}
\end{subequations}
The strength of the coupling parameter $g_{0}=g_{mb}+g_{pe}$ between two ($m$-mechanical, $n$-optical) eigenmodes strongly depends on the overlap of the displacement $u_{m}(z)$ or strain $\frac{\partial u_{m}}{\partial z}(z)$ field with the electric field intensity $|E_{n}(z)|^2$,
which highlights the importance of co-localizing the displacement and the electric field profiles.\ Since our model is by definition one-dimensional, we have chosen an \textit{arbitrary} size in the x-y plane of $2\times 2$ $\mu m^{2}$ -for the area excited by a focused laser beam- in order to calculate the effective mass $m_{eff}$ of the mechanical resonator and the zero point fluctuations $x_{zpf} =\sqrt{\frac{\hbar}{2m_{eff}\Omega_{m}}}$.\ To quantify the role of co-localization in the optomechanical coupling, we will evaluate $g_{0}$ in the GaAs/AlAs alloy as well as in a Si/Ge supperlattice where the condition (\ref{conditions}) is not fulfilled.\\

To calculate the terms \ref{gmb} and \ref{gpe}, we apply the transfer matrix method with outgoing boundary conditions which define quasi-normal modes~\cite{qnormalmodes} with a complex eigenfrequency $\omega$ = $\omega_{r}$ - $j\omega_{i}$ (see details in the supplementary information).\ The imaginary part accounts for losses through the walls of the resonator, uppermost GaAs layer and bottom most AlAs layer in the superlattice, which is equivalent to using a Perfectly Matched Layer (PML) in numerical simulations (FEM, FDTD, etc.) on three dimensional structures~\cite{PML}.\ In particular, we use the standard complex root-finding M\"{u}ller method to solve the fundamental equation, calculating the initial values from the central frequency and the linewidth of the transmission resonances.\ When the method failed to converge, usually due to closely spaced eigenvalues, we used a method based on the argument principle method (APM) of complex analysis~\cite{APM}.\ The boundary conditions assumed correspond to a semi-infinite substrate and a free-moving surface.\ For a large number of layers, however, the boundary conditions have essentially negligible effect precisely deep in the localization regime.\ Based on these arguments, we use the resonances of the mechanical transmission spectra to extract initial values for the real and imaginary part of the eigenfrequencies.\\

\begin{figure}[t]
  \includegraphics[width=\columnwidth]{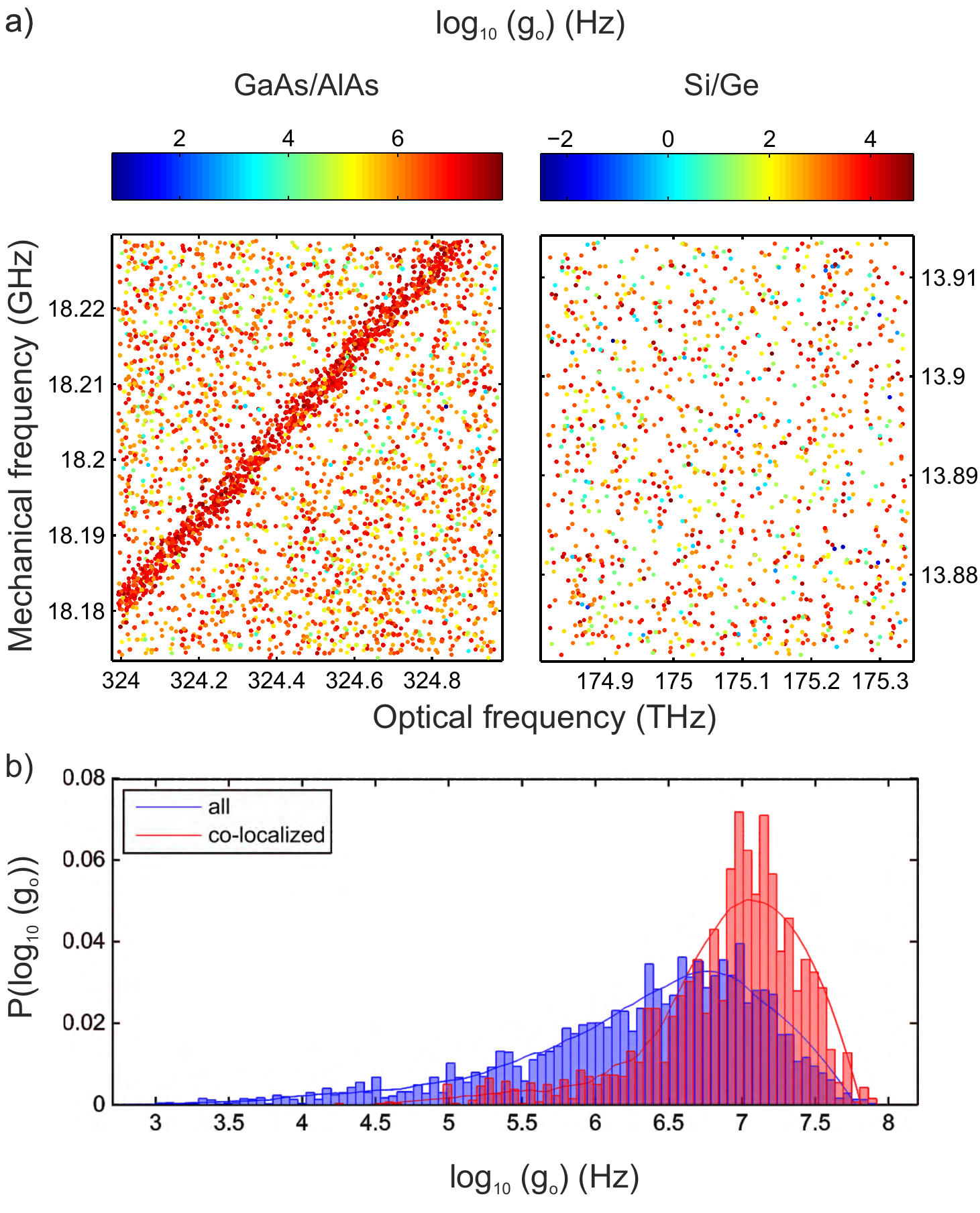}
    \caption{ \label{3} \textbf{Vacuum optomechanical coupling rate $g_{0}$ in disordered multilayered systems.} (a) Scatter plot of the mechanical and optical frequencies of the resonant modes found in a
    set of 1500 disordered GaAs/AlAs (left) and Si/Ge (right) multilayers, the color represents the coupling rate $g_{0}$ between the considered pair in logarithmic scale.\ (b) Probability density function of the coupling rate $g_{0}$
    extracted from the left panel in (a) by considering all photon-phonon pairs (blue) and only perfectly co-localized pairs (red).\ The thick lines are only guides to the eye. }
\end{figure}

Finally, we calculate the vacuum optomechanical coupling rate $g_{0}$ - Fig.~\ref{3} - between all the photonic and phononic Anderson-localized modes found in a set of 1500 GaAs/AlAs superlattices with the same structural parameters as detailed above and a fixed level of disorder of $\sigma=0.15\cdot D$.\ The coupling values in the densely-packed \textit{diagonal} of Fig.~\ref{3}(a), i.e., for perfectly co-localized photon-phonon pairs, exhibit values higher than the rest.\ The probability density function of the vacuum optomechanical coupling rate $g_{0}$ is plotted in Fig.~\ref{3}(c) either considering all eigenmode pairs (blue bars) or only those \textit{perfectly} co-localized (red bars), pointing out a marked statistical evidence for this material combination.\ Due to the strongly dispersive nature of the localization length $\xi$ (Fig~\ref{1}(b)), we consider only a narrow frequency window thus minimizing the effect of varying the effective volume of the eigenmodes.\ To point out explicitly the unique character of the combination of GaAs and AlAs to engineer the co-localization of mechanical and optical modes, we apply the same calculation to Si/Ge disordered superlattices.\ The Si and Ge thicknesses ($d_{1}=115.97$ nm, $d_{2}=93.88$ nm, $N=600$) are chosen to shift the band edge of interest at approximately half the frequency of the GaAs/AlAs superlattice to avoid the effect of absorption.\ The disorder level is also tuned to $\sigma=0.11\cdot D$ to satisfy the condition for the mechanical localization length $\xi_{Si/Ge}\sim\xi_{GaAs/AlAs}$.\ Under these conditions, the Si/Ge superlattices show no predictable modal structure as plotted in Fig.~\ref{3}(b).

In conclusion, we present an analysis of Anderson co-localization of two wave fields in a particular set of random superlattices, with emphasis in their coupling properties.\ Just by selecting GaAs and AlAs as the superlattice materials, it is possible to enhance their optomechanical coupling and allows further optimization trough field engineering while relaxes the design and fabrication accuracy.\ Our calculations also demonstrate the potential of such a system to explore optomechanical coupling in the Anderson-localization regime by providing quantitative evidence of the role played by spatial co-localization of mechanical and optical excitations in the structure.\ Light-matter interaction in GaAs/AlAs DBR-based cavity structures appears to be a natural and appropriate choice to explore Anderson localization of coupled excitations and to observe Anderson localization of phonons in frequency ranges so far unexplored.\ The ease of integration of quantum wells and quantum dots~\cite{LasingCzerniuk} during the MBE growth enables the study of cavity quantum-electrodynamics in the Anderson-localization regime coupled to the mechanical motion of a high frequency nanomechanical oscillator, as well as to explore the role of phonon-coupling in the performance of random lasers~\cite{Jin}.\ Finally, the discussed structures can be easily scaled down to study Anderson localization of sub-THz vibrations with extended light modes.

This work was supported by the Spanish MINECO via the Severo Ochoa Program (Grant SEV-2013-0295) and the project PHENTOM (Fis 2015-70862-P), as well as by  the CERCA Programme / Generalitat de Catalunya,
and by the European Commission in the form of the H2020 FET Open project PHENOMEN (GA. Nr. 713450).\ GA is supported by a BIST PhD. Fellowship, NDLK by the ERC grant nr. 715939 and PDG by a Ramon y Cajal
fellowship nr. RyC-2015-18124.


\newpage

\section{supplementary information}

In our calculations, we use transfer matrix formalism~\cite{rytovmodel}.\ In particular, we consider time-harmonic longitudinal acoustic waves with no in-plane dependence, i.e., $\mathbf{u}(\mathbf{r},t) \equiv u_{z}(z)e^{j\omega t}\mathbf{e}_{z}$, where $\omega$ is the mechanical frequency.\ The solution to the continuum-mechanics equations in the absence of sources in the bulk of layer $i$ can be written as a linear combination of a propagating and counter-propagating plane wave, i.e., $u_{z}(z)=a_{i}e^{jq_{i}z}+b_{i}e^{-jq_{i}z}$, with the wave-vector $q_{i}=\frac{\omega}{v_{i}}$.\ Imposing continuity on both the displacement field and the stress at the interfaces and accounting for propagation within the layers, the coefficients in the plane wave basis on the incident layer, here the left side of the superlattice, i.e. the first solid interface, $(a_{0}^d,b_{o}^d)$ are algebraically connected to the solution on the right side of the last layer, i.e. top of the substrate, $(a_{N+1},b_{n+1})$ as:
\begin{equation}
    \begin{bmatrix}
        a_{0}^d \\
        b_{0}^d
    \end{bmatrix}
    =
    \mathbf{M}_{ac} \boldsymbol{\cdot}
    \begin{bmatrix}
        a_{N+1} \\
        b_{N+1}
    \end{bmatrix}
    \quad \text{with}\quad \mathbf{M}_{ac}=(\prod_{i}^{N-1} \mathbf{I}_{i,i+1}\cdot\mathbf{Y}_{i+1})\cdot \mathbf{I}_{N,N+1}
\end{equation}
where the subindex \emph{ac} in the matrix $\mathbf{M}_{ac}$ stands for acoustic.\ The $\mathbf{I}_{i,i+1}$ is an interface matrix connecting the coefficients in the propagating/counter-propagating plane wave basis at the end of layer $i$ to the ones at the beginning of layer $i+1$
and $\mathbf{Y}_{i}$ propagates the solution through the thickness of layer $i$:
\begin{subequations}\label{matrices}
\begin{align}
  \mathbf{I}_{i,i+1}  &=
    \begin{bmatrix}
        1 + \frac{Z_{i+1}}{Z_{i}} & 1 - \frac{Z_{i+1}}{Z_{i}} \\
        1 - \frac{Z_{i+1}}{Z_{i}} & 1 + \frac{Z_{i+1}}{Z_{i}}
    \end{bmatrix} \label{matrixa}\\
    \mathbf{Y}_{i} &=
    \begin{bmatrix}
        e^{-jq_{i}d_{i}} &  0 \\
        0 & e^{+jq_{i}d_{i}}
    \end{bmatrix} \label{matrixb}
\end{align}
\end{subequations}

Interestingly, Maxwell's equations at normal incidence in such a multilayer structure can also be solved with exactly the same formalism~\cite{TMM}, provided that the acoustic impedances $Z_{i}$ are replaced by refractive indexes $n_{i}$ and the longitudinal acoustic velocities $v_{i}$ by light speeds $\frac{c}{n_{i}}$ in matrices (\ref{matrixa}) and (\ref{matrixb}).\ A direct consequence of the equivalence in the formalism is:

\begin{equation}
\label{conditions}
 \left \{
   \begin{array}{l}
      \frac{Z_{i+1}}{Z_{i}} = \frac{n_{i+1}}{n_{i}}  \\
      \frac{c/n_{i}}{v_{i}} = K
   \end{array} \right\}  \forall i \in [0,N+1] \implies  \mathbf{M}_{ac}(\omega)=\mathbf{M}_{op}(K\omega)
\end{equation}

For any arbitrary multilayered structure and, in particular, any combination of the two materials (A and B) forming a superlattice, condition (\ref{conditions}) is extremely stringent. The contrast between their acoustic impedances ($Z_{A}/Z_{B}$) and their refractive indexes ($n_{A}/n_{B}$) is not equal, neither the ratio of the propagation speeds ($\frac{c/n_{A/B}}{v_{A/B}}$) stays constant.\ The optical matrix $\mathbf{M}_{op}$ is, therefore, different than the acoustic one which gives rise to independent acoustic and optical interference processes.\ This leads to an unavoidably different mechanical and optical field profiles where co-localization occurs very rarely.\ Remarkably, for GaAs and AlAs, in non-dispersive frequency regions, this condition is approximately satisfied, which is the mechanism to guarantee an almost perfect spatial co-localization of the electromagnetic and the mechanical fields.\\

The use of different types of boundary conditions for the physical problem in hand gives rise to different applications of the transfer matrix method:

\begin{itemize}
\item For transmission/reflection computations, the problem is set with $a_{0}^{d}=1$, $b_{0}^{d}=r$, $a_{N+1}=t$, $b_{N+1}=0$, which implies $r=\frac{M_{21}}{M_{11}}$ and $t=\frac{1}{M_{11}}$, from which the reflectivity and transmissivity of the superlattice can be calculated for any frequency $\omega$.
\item For the optical eigenmodes, since the structure is inherently an open structure at both ends, we set $a_{0}^{d}=0$ and $b_{N+1}=0$, i.e. outgoing boundary conditions. These two conditions necessarily imply $M_{11}(\omega)=0$, which is solved for complex eigenfrequencies $\omega$ = $\omega_{r}$ - $j\omega_{i}$.
\item For the mechanical eigenmodes we assume a closed structure on the top interface and an open structure at the bottom, i.e. a semi-infinite substrate. A much thicker substrate thickness than the structures of interest and unavoidable sound absorption justify such choice for the bottom interface. The stress-free boundary condition ($\mathbf{\sigma\cdot n}$) on the top interface implies $a_{0}^{d}=1$ and $b_{0}^{d}=0$, while the outgoing boundary condition at the bottom implies $b_{N+1}=0$. These conditions imply $M_{21}(\Omega)-M_{21}(\Omega)=0$, which is solved for complex eigenfrequencies $\Omega$ = $\Omega_{r}$ - $j\Omega_{i}$.
\end{itemize}

\end{document}